\documentclass[%
preprint,
amsmath,amssymb,
aps,
prx,
]{revtex4-1}
\usepackage{pgfplots}
\usepackage{amsmath, amsthm, amssymb, amsfonts}

\usepackage[colorlinks=true]{hyperref}  
\hypersetup{
	bookmarks=true,         % show bookmarks bar?
	unicode=false,          % non-Latin characters 
	pdftoolbar=true,        % show Acrobat
	pdfmenubar=true,        % show Acrobat 
	pdffitwindow=false,     % window fit to page when opened
	pdfstartview={FitH},    % fits the width of the page to the window
	pdftitle={Fermi surface reconstruction without symmetry breaking},    % title
	pdfauthor={Snir Gazit and Subir Sachdev},     % author
	pdfsubject={},   % subject of the document
	pdfcreator={},   % creator of the document
	pdfproducer={}, % producer of the document
	pdfkeywords={} {} {}, % list of keywords
	pdfnewwindow=true,      % links in new window
	colorlinks=true,       % false: boxed links; true: colored links
	linkcolor=magenta, %red, % color of internal links (change box color with linkbordercolor)
	citecolor=blue,        % color of links to bibliography
	filecolor=magenta,      % color of file links
	urlcolor=blue           % color of external links
} 
\usepackage{cleveref}
\usepackage{xcolor}
\usepackage{fullpage}
\usepackage{amsmath}
\usepackage{amsfonts}
\usepackage{braket}
\usepackage{dsfont}
\usepackage{graphicx}

\usepackage{subcaption}

\newcommand{\av}[1]{\left\langle #1\right\rangle}
\usepackage{amsmath}
\DeclareMathOperator{\Tr}{\text{Tr}}
\DeclareMathOperator{\Imag}{\text{Im}}
\newcommand{\ztwo}{\mathbb{Z}_2}

\newcommand{\subfp}[1]{\begin{subfigure}[b]{0.4\textwidth}
		\phantomcaption
		\label{#1}
\end{subfigure}}     
\begin{document}

	\title{Fermi surface reconstruction without symmetry breaking}
	\author{Snir Gazit}
	\affiliation{Racah Institute of Physics and The Fritz Haber Research Center for Molecular Dynamics, The Hebrew University of Jerusalem, Jerusalem 91904, Israel}
	\author{Fakher F. Assaad}
	
	\affiliation{Institut f\"{u}r Theoretische Physik und Astrophysik,
		Universit\"{a}t W\"{u}rzburg, Am Hubland, D-97074 W\"{u}rzburg, Germany}
	\affiliation{W\"{u}rzburg-Dresden Cluster of Excellence ct.qmat,
		Universit\"{a}t W\"{u}rzburg, Am Hubland, D-97074 W\"{u}rzburg, Germany
	}
	\author{Subir Sachdev}
	\affiliation{Department of Physics, Harvard University, Cambridge MA 02138, USA}
	\begin{abstract}
		We present a sign-problem free quantum Monte Carlo study of a model that exhibits quantum phase transitions without symmetry breaking and associated changes in the size of the Fermi surface. The model is an Ising gauge theory on the square lattice coupled to an Ising matter field and spinful `orthogonal' fermions at half-filling, both carrying Ising gauge charges. In contrast to previous studies, our model hosts an electron-like, gauge-neutral fermion excitation providing access to Fermi liquid phases. One of the phases of the model is a previously studied orthogonal semi-metal, which has $\mathbb{Z}_2$ topological order, and Luttinger-volume violating Fermi points with gapless orthogonal fermion excitations. We elucidate the global phase diagram of the model: along with a conventional Fermi liquid phase with a large Luttinger-volume Fermi surface, we also find a 
			`deconfined' Fermi liquid in which the large Fermi surface co-exists with fractionalized excitations. We present results for the electron spectral function, showing its evolution from the orthogonal semi-metal with a spectral weight near momenta $\{\pm \pi/2, \pm \pi/2\}$, to a large Fermi surface.
	\end{abstract}
	\maketitle
	
	\section{Introduction}
	\label{sec:intro}
	
	Quantum phase transitions involving a change in the volume enclosed by the Fermi surface play a fundamental role in correlated electron compounds. 
	In the cuprates, there is increasing evidence of a phase transition from a low doping pseudogap metal state with small density of fermionic quasiparticles, to a higher doping Fermi liquid (FL) state with a large Fermi surface of electronic quasiparticles \cite{Keimer2015,PTARCMP,Fujita14,He14,badoux_change_2016,MichonNature,Ramshaw2020,Sachdev_2018,Tsvelik19}. In the heavy fermion compounds, much attention has focused on the transitions between metallic states distinguished by whether the Fermi volume counts the localized electronic $f$ moments or not \cite{Stewart01,Qimiao2010,SVS04,Coleman01}. 
	
	Given the strong coupling nature of such transitions, quantum Monte Carlo simulations can offer valuable guides
	to understanding the consequences for experimental observations. As the transitions involve fermions at non-zero density, the sign problem is a strong impediment to simulating large systems. However, progress has been possible in recent years by a judicious choice of microscopic Hamiltonians which are argued to capture the universal properties of the transition but are nevertheless free of the sign problem. Such approaches have focused on density wave ordering transitions \cite{BergARCMP,BMS12,WangLee1,WangLee2,Schattner16,Huffman:2017swn,Gerlach17,Rafael17,Rafael18,Sato18,Meng1,Meng3,Meng4,YaoARCMP}, where spontaneous translational symmetry breaking accompanies the change in the Fermi volume: consequently both sides of the transition have a Luttinger volume Fermi surface, after the expansion of the unit cell by the density wave ordering has been taken into account. 
	
	Our paper will present Monte Carlo results for quantum phase transitions without symmetry breaking, accompanied by a change in the Fermi surface size from a non-Luttinger volume to a Luttinger volume. 
	The phase with a non-Luttinger volume Fermi surface must necessarily have topological order and emergent gauge degrees of freedom \cite{SVS04,Paramekanti_2004,Ersatz2020}.
	There have been a few quantum Monte Carlo studies of fermions coupled to emergent gauge fields \cite{GazitZ2,Grover16,Gazit_pnas,Hofmann_2018,Assaad18,Wu16,Meng5,Meng2}, and
	our results are based on a generalization of the model of refs.~\cite{GazitZ2,Gazit_pnas}, containing a $\mathbb{Z}_2$ gauge field coupled to spinful `orthogonal' fermions $f_\alpha$ (spin index $\alpha=\uparrow, \downarrow $) carrying a $\mathbb{Z}_2$ gauge charge hopping on the square lattice at half-filling. The parameters are chosen so that the $f_\alpha$ experience an average $\pi$-flux around each plaquette; consequently when the fluctuations of the $\mathbb{Z}_2$ gauge flux are suppressed in the deconfined phase of the $\mathbb{Z}_2$ gauge theory, the fermions have the spectrum of massless Dirac fermions at low energies. There are four species of two-component Dirac fermions, arising from the two-fold degeneracies of spin and valley each. In addition to the $\mathbb{Z}_2$ gauge charges, the $f_\alpha$ fermions also carry spin and global U(1) charge quantum numbers, and so are identified with the orthogonal fermions of Ref.~\cite{Nandkishore_2012}, and the deconfined phase of the $\mathbb{Z}_2$ gauge theory is identified as an orthogonal semi-metal (OSM). 
	\begin{figure}
		\centering
		\begin{tabular}{cc}
			\begin{subfigure}[t]{0.5\textwidth}\centering\caption{}
				\includegraphics[scale=0.55]{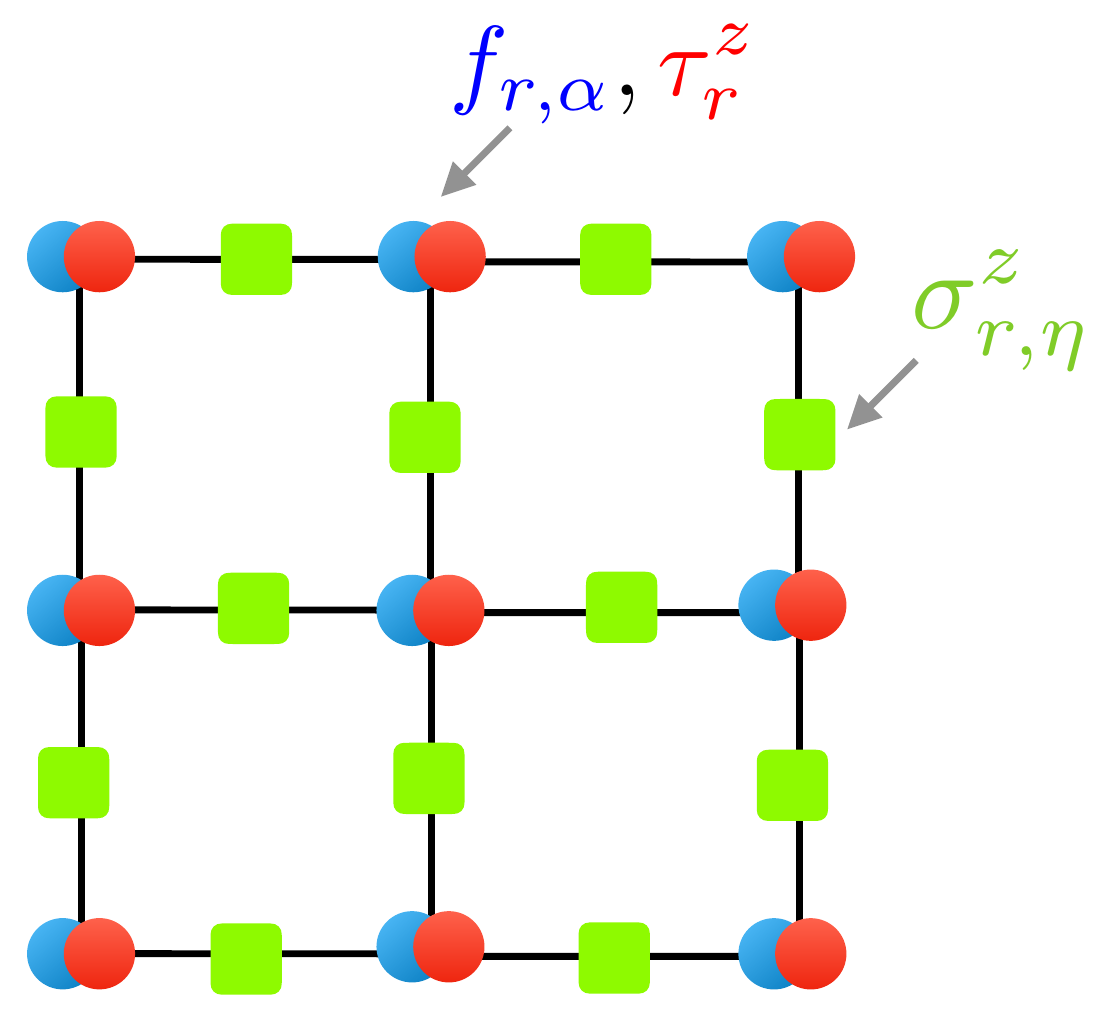}
				\label{subfig:model}
			\end{subfigure} &
			\begin{subfigure}[t]{0.5\textwidth}\centering\caption{}
				\includegraphics[scale=0.95]{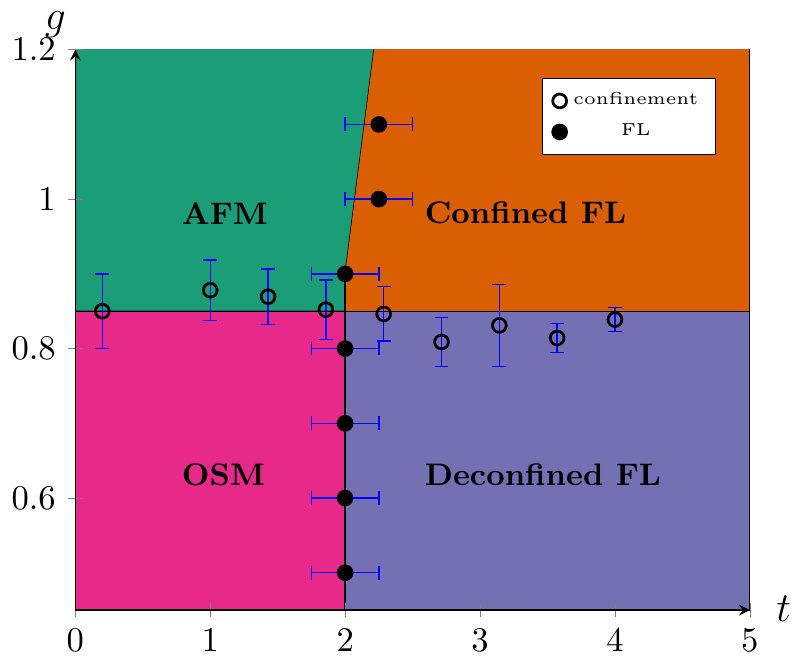}
				\label{subfig:phasediag}
			\end{subfigure}
			%\subf{\input{Figs/phase_diag.tex}} 
		\end{tabular}
		\centering
		\caption{(a) Lattice model of orthogonal fermions coupled to an Ising-Higgs lattice gauge theory. The matter fields $f_{r,\alpha}$ (blue circle) and $\tau^z_r$ (red circle), reside on the square lattice sites and the Ising gauge field $\sigma^z_{r,\eta}$ (green square) is defined on the lattice bonds. (b) Global phase diagram of our model (\cref{eq:H_ising,eq:Hf}) as a function of the hopping amplitude, $t$, and transverse field, $g$. The phase boundaries are determined by the location of the confinement transition and emergence of $c$ fermions spectral weight, see main text. Connecting lines are guide to the eye. }
	\end{figure}
	
	It is important to note that the OSM phase preserves all the symmetries of the square lattice Hamiltonian, and it realizes a phase of matter which does {\it not\/} have Luttinger volume Fermi surface. This is compatible with the topological non-perturbative formulation of the Luttinger theorem (LT) \cite{Oshikawa_2000} because $\mathbb{Z}_2$ flux has been expelled (about a $\pi$-flux background), and the OSM has $\mathbb{Z}_2$ topological order \cite{SVS04,Paramekanti_2004,Ersatz2020}. 
	There is a Luttinger constraint associated with every unbroken global U(1) symmetry \cite{Powell05,Coleman05}, stating that the total volume enclosed by Fermi surfaces of quasiparticles carrying the global charge (along with a phase space factor of $1/(2 \pi)^d$, where $d$ is spatial dimension) must equal the density of the U(1) charge, modulo filled bands. 
	In Oshikawa's argument \cite{Oshikawa_2000}, this constraint is established by placing the system on a torus, and examining the momentum balance upon insertion of one quantum of the global U(1) flux: the Luttinger result follows with the {\it assumption\/} that the only low energy excitations which respond to the flux insertion are the quasiparticles near the Fermi surface. When there is $\mathbb{Z}_2$ topological order, a $\mathbb{Z}_2$ flux excitation (a `vison') inserted in the cycle of the torus costs negligible energy and can contribute to the momentum balance: consequently, a non-Luttinger volume Fermi surface becomes possible (but is not required) in the presence of topological order \cite{SVS04,Paramekanti_2004,Ersatz2020}. In the OSM, the orthogonal fermions $f_\alpha$ carry a global U(1) charge and have a total density of 1: so in the conventional Luttinger approach, there must be 2 Fermi surfaces (one per spin) each enclosing volume $(1/2) (2 \pi)^2$. However, with $\mathbb{Z}_2$ topological order, the OSM can evade this constraint, and the only zero energy fermionic excitations are at discrete Dirac nodes, and the Fermi surface volume is zero. Earlier numerical studies \cite{GazitZ2,Grover16,Gazit_pnas} presented indirect evidence for the existence of such a Luttinger-violating OSM phase, and we will present direct evidence here in spectral functions.
	
	Our interest here is in quantum transitions out of the OSM, and in particular, into phases without $\mathbb{Z}_2$ topological order. In the previous studies \cite{GazitZ2,Grover16,Gazit_pnas}, the $\mathbb{Z}_2$ confined phases broke either the translational or the global U(1) symmetry. In both cases, there was no requirement for a Fermi surface with a non-zero volume, and all fermionic excitations were gapped once $\mathbb{Z}_2$ topological order disappeared. Here, we will extend the previous studies by including an Ising matter (Higgs) field, $\tau^z$, which also carries a $\mathbb{Z}_2$ gauge charge. This allows us to define a gauge-invariant local operator with the quantum number of the electron \cite{Nandkishore_2012}:
	\begin{equation}
	c_\alpha = \tau^z f_\alpha
	\label{eq:OF}
	\end{equation}
	The $\tau^z$ matter fields do not carry global spin or U(1) charges. With this dynamic Ising matter field present, it is possible to have a $\mathbb{Z}_2$-confined phase which does not break any symmetries, and phase transitions which are not associated with broken symmetries. In particular, the Luttinger constraint implies that any $\mathbb{Z}_2$ confined phase without broken symmetries must have large Fermi surfaces with volume $(1/2) (2 \pi)^2$ for each spin. Our main new results are measurements of the $c_\alpha$ spectral function with evidence for such phases and phase transitions.
	
	One of the unexpected results of our Monte Carlo study is the appearance of an additional ``Deconfined FL'' phase: see the phase diagram in \cref{subfig:phasediag}. As we turn up the attractive force between the $f_\alpha$ fermions and the Ising matter (Higgs) field $\tau^z$, we find a transition from the OSM to a phase with a large Luttinger-volume Fermi surface of the $c_\alpha$, but with the $\mathbb{Z}_2$ gauge sector remaining deconfined.  Such a phase does not contradict the topological arguments, which do not require (but do allow) a Luttinger-violating Fermi surface in the deconfined state. Only when we also turn up the $\mathbb{Z}_2$ gauge fluctuations do we then get a transition to a ``Confined FL'' phase, which is a conventional Fermi liquid with a large Fermi surface. Within the resolution of our current simulations, we were not able to identify a direct transition from the OSM to the Confined FL, and we have indicated a multicritical point separating them in \cref{subfig:phasediag}.
	
	It is interesting to note that the Deconfined FL phase has some features in common with `fractionalized Fermi liquid' (FL*) phases used in recent work \cite{Punk_2015,Sachdev_2018,Sachdev:2018nbk} to model the pseudogap phase of the cuprates.
	These phases share the presence of excitations with deconfined $\mathbb{Z}_2$ gauge charges co-existing with a Fermi surface of gauge-neutral fermions $c_\alpha$. There is a difference, however, in that the present Deconfined FL state has a large Fermi surface, while the FL* states studied earlier had a small Fermi surface. Both possibilities are allowed by the topological LT. \cite{SVS04,Paramekanti_2004}.
	
	In passing, we note the recent study of Chen {\it et al.\/} \cite{Meng2}, which also examined a $\mathbb{Z}_2$ gauge theory coupled to orthogonal fermions, $f_\alpha$, and an Ising matter field, $\tau^z$. However, their $\mathbb{Z}_2$ deconfined phase is different from ours and earlier studies \cite{Grover16,Gazit_pnas}: their phase has a large, Luttinger-volume Fermi surface of the $f_\alpha$ fermions which move in a background of zero flux (such zero flux states were also present in the studies of ref.~\cite{GazitZ2,Assaad18}). Consequently, their $\mathbb{Z}_2$ confinement transition to the Fermi liquid does not involve a change in the Fermi surface volume, and this weakens the connection to finite doping transitions in the cuprate superconductors. 
	
	The rest of the paper is structured as follows. In \cref{sec:model} we introduce a lattice realization of the Ising-Higgs gauge theory coupled to orthogonal fermions and discuss its global and local symmetries, in \cref{sec:qmc} we determine the global phase diagram of our model using  a sign-problem-free QMC simulation. In particular, we study the structure of the Fermi surface and state of the gauge sector in the different phases, and comment on the nature of the numerically observed quantum phase transitions, in \cref{sec:mf} we present a mean field calculation of the physical fermion spectral function in the OSM phase, and lastly in \cref{sec:summary} we summarize our results, discuss relations to experiments and highlight future directions. 
	
	\section{Ising-Higgs gauge theory coupled to orthogonal fermions}
	\label{sec:model}
	
	\subsection{Lattice Model}

	As a concrete microscopic model for orthogonal fermions, we consider the square lattice model depicted in \cref{subfig:model}. The dynamical degrees of freedom are Ising gauge fields, $\sigma^z_{b}=\pm1$, residing on the square lattice bonds $b=\{r,\eta\}$, with $r$ being the lattice site and $\eta=\hat{x}/\hat{y}$, and two types of matter fields: an Ising field $\tau^z_{r}=\pm1$ and a spinful orthogonal fermion $f_{\alpha,r}$, with $\alpha=\uparrow,\downarrow$ labeling the spin index. Both matter fields are defined on the lattice sites.  The dynamics is governed by the Hamiltonian, $\mathcal{H}=\mathcal{H}_{\ztwo}+\mathcal{H}_{\tau}+\mathcal{H}_f+\mathcal{H}_c$ comprising the lowest order terms that are invariant under local Ising gauge transformations, as we detail below. 
	
	The first two terms in $\mathcal{H}$ correspond to the standard Ising-Higgs gauge theory \cite{fradkin_2013},
	\begin{equation}
	\begin{split}
	\label{eq:H_ising}
	\mathcal{H}_{\ztwo}&=-K\sum_{\square} \prod_{b\in\square} \sigma^z_b-g \sum_b \sigma^x_b  \\ 
	\mathcal{H}_{\tau}&=-J\sum_{r,\eta} \sigma^z_{r,\eta}\tau^z_r \tau^z_{r+\eta} -h\sum_r \tau^x_r.
	\end{split}
	\end{equation}
	In the above equations, the operators $\boldsymbol{\sigma}=\{\sigma^x,\sigma^y,\sigma^z\}$ and  $\boldsymbol{\tau}=\{\tau^x,\tau^y,\tau^z\}$ are the conventional Pauli matrices, acting on the Hilbert spaces of the Ising gauge fields and Ising matter fields, respectively. The Ising magnetic flux term ,$\Phi_\square=\prod_{b\in\square} \sigma_b^z$, in $\mathcal{H}_{\ztwo}$ equals to the product of the Ising gauge field belonging to the elementary square lattice plaquettes, $\square$. $\mathcal{H}_{\tau}$ is a transverse field Ising Hamiltonian for the Ising matter field, where in order to comply with Ising gauge invariance the standard Ising interaction is modified to include an Ising gauge field, $\sigma^z_{b}$, along the corresponding bonds $b$.
	
	The fermion dynamics is captured by the last two terms in $\mathcal{H}$,
	\begin{equation}
	\begin{split}
	\mathcal{H}_{f}&=-w\sum_{r,\eta,\alpha} \sigma^z_{r,\eta}f^\dagger_{r,\alpha} f_{r+\eta,\alpha}+h.c+U\sum_r \left(n^f_{r,\uparrow}-\frac{1}{2}\right)\left(n^f_{r,\downarrow}-\frac{1}{2}\right)\\ 
	\mathcal{H}_{c}&=-t\sum_{r,\eta,\alpha} \tau^z_{r}f^\dagger_{r,\alpha} \tau^z_{r+\eta}f_{r+\eta,\alpha}+h.c.
	\end{split}
	\label{eq:Hf}
	\end{equation}
	Here, $\mathcal{H}_f$ includes a gauge invariant nearest-neighbor hopping of orthogonal fermions and on-site Hubbard interaction between fermion densities, $n^f_{r,\alpha}=f^\dagger_{r,\alpha}f_{r,\alpha}$. The last term, $\mathcal{H}_c$, defines nearest-neighbor hopping of physical (gauge-neutral) $c_{r,\alpha}$ fermions as can be readily verified by substituting \cref{eq:OF}. The model is tuned to half-filling (the chemical potential vanishes).
	
	In relation to past works, the model considered here affords a non-trivial generalization of the ones studied previously in refs. \cite{GazitZ2,Grover16,Gazit_pnas}. In particular, in contrast to prior studies, where the Ising matter fields $\tau^z$ were infinitely massive ($h\to\infty$ in \cref{eq:H_ising}), here, varying the transverse field, $h$, in \cref{eq:H_ising} controls the excitation gap for $\tau^z$ particles. Consequently, the Ising matter fields and subsequently the physical (gauge-neutral) fermion $c_{\alpha}=f_{\alpha} \tau^z$ are now dynamical degrees of freedom. This important extension provides access to a more generic phase diagram and observables that probe FL physics.
	
	\subsection{Global and local symmetries}
	
	We now turn to discuss the global and local symmetries of our model. $\mathcal{H}$ is invariant under a global $SU_s(2)$ symmetry of spin rotations. Furthermore, because we tune to half filling, particle-hole symmetry enlarges the $U(1)$ symmetry, corresponding to fermion number conservation, to form a $SU_c(2)$ pseudo-spin symmetry \cite{auerbach_book}. Physically, the $SU_c(2)$ symmetry generates rotations between the charge density wave and s-wave superconductivity order parameters.
	
	The gauge structure of our model is manifest through the invariance of $\mathcal{H}$ under an infinite set of {\em local} $\ztwo$ gauge transformations generated by the operators $G_r=(-1)^{n^f_r}\tau^x_r\prod_{b\in +_r} \sigma^x_b $. Here, $n^f_r=\sum_\alpha n^f_{r,\alpha}$ and $+_r$ denotes the set of bonds emanating from the site $r$. Because $[\mathcal{H},G_r]=0$ for {\it all} sites $r$, the eigenvalue $Q_r=\pm1$ of $G_r$ are conserved quantities. Physically, $Q_r$ is identified with the static on-site $\ztwo$ background charge assignment. 
	
	To properly define a gauge theory, one must fix the background charge configuration, $Q_r$. This procedure enforces an Ising variant of Gauss's law $G_r=Q_r$. Requiring a translationally invariant configuration, two distinct gauge theories may be defined: an even lattice gauge theory with a trivial background ($Q_r=1$) and an odd lattice gauge theory with a single Ising charge at each site ($Q_r=-1$). For concreteness, in what follows, we will only consider the case of an odd lattice gauge theory. As explained in ref.~\cite{Gazit_pnas}, at half-filling, the corresponding results for the even sector may be obtained by applying a partial particle-hole transformation acting on one of the spin species \cite{auerbach_book}.
	
	Our model is also invariant under discrete square lattice translations. The operators $\hat{T}_{x}$ and $\hat{T}_y$ generate translation by a lattice constant along the $x$ and $y$ directions, respectively. When acting on fractionalized excitations, such as the matter fields $\tau^z_r$ and $f_{r,\alpha}$ in our case, translations may be followed by a $\ztwo$ gauge transformation. The symmetry operation then forms a {\it projective} representation. In the general case, this allows for a richer group structure than the standard (gauge-neutral) linear representation \cite{Wen_book}.
	
	For the specific case of a $\ztwo$ gauge symmetry on a square lattice, a projective implementation of translations is potentially non-trivial. In particular, lattice translation along the $x$ and $y$ directions may either commute or anti-commute \cite{Wen_book}, namely $T_x T_y=\pm T_y T_x$. Physically, the former corresponds to trivial translations, whereas the latter defines a $\pi$-flux pattern threading each elementary plaquettes of the square lattice. While for every given choice of a gauge fixing condition, the $\pi$-flux lattice inevitably breaks lattice translations, as it leads to doubling of the unit cell, for fractionalized excitations, translational symmetry is restored by applying an Ising gauge transformation \cite{Wen_book}. This key observation allows for the OSM phase to violate LT without breaking of translational symmetry.
	
	\section{Quantum Monte Carlo simulations}
	\label{sec:qmc}
	\subsection{Methods}
	
	Our model is free of the numerical sign-problem. We can, therefore, elucidate its phase diagram using an unbiased and numerically exact (up to statistical errors) QMC calculations. To control the Trotter discretization errors, we set the imaginary time step to satisfy $\Delta \tau\le1/(12|t|)$, a value for which we found that discretization errors are sufficiently small to obtain convergent results. We explicitly enforce the Ising Gauss law using the methods introduced in ref.~\cite{GazitZ2,Gazit_pnas}. Additional details discussing the implementation of the auxiliary-field QMC algorithm and its associated imaginary time path-integral formulation are given in \cref{app:path_integral}. Similar results were obtained without imposing the constraint and using the and using the algorithms for lattice fermions (ALF) library  \cite{ALF_v1}.
	
	\subsection{Observables}
	
	To track the evolution of the $c$ electron Fermi surface, we will study the imaginary-time two point Green's function, $\mathcal{G}^\beta_\alpha(k,\tau)=-\av{\mathcal{T} [c_{k,\alpha}(\tau)c_{k,\alpha}^\dagger(0)]}$, where $\mathcal{T}$ denotes time ordering and the operator $c^\dagger_{k,\alpha}=\sum_r f^\dagger_{r,\alpha} \tau^z_re^{ik\cdot r}$ creates a $c$ fermion carrying momentum $k$ and spin polarization $\alpha$. Expectation values are taken with respect to the thermal density matrix, $\av{O}=\frac{1}{\mathcal{Z}}\Tr\left[e^{-\beta \mathcal{H}}O\right]$, with $\mathcal{Z}=\Tr\left[e^{-\beta \mathcal{H}}\right]$ being the thermal partition function at inverse temperature $\beta=1/T$. We emphasize that in contrast to the orthogonal $f$ electron, for which, in the absence of a string operator, gauge invariance requires that the two-point Green's function must vanish for all non-equal space-time points, the $c$ electron is a gauge-neutral operator and hence its associated spectral function may be non-trivial.
	
	Determining the Fermi surface structure requires knowledge of real time quantum dynamics. Therefore, some form of analytic continuation of the imaginary time QMC data to real frequency must be carried out. Quite generically, devising a reliable and controlled numerical analytic continuation technique is an outstanding challenge due to the inherent instability of the associated inversion problem \cite{Gazit_AC}.
	
	To overcome this difficulty, we employ a commonly used proxy for the low frequency spectral response \cite{Nandini_Gtau}. More explicitly, by computing $G^\beta(k,\tau)$ (the spin index was omitted for brevity) at the largest accessible imaginary time difference $\tau=\beta/2$, we obtain an estimate for the single particle residue $Z$. To see how to relate this quantity to real time dynamics, we consider the integral relation 
	\begin{equation}
	\mathcal{G}^\beta(k,\tau=\beta/2)=-\int_{-\infty}^\infty d \omega \,\frac {A(k,\omega)}{2\cosh(\beta \omega)},
	\end{equation}
	where $A^\beta(k,\omega)=-\frac{1}{\pi}\Imag \mathcal{G}^\beta(k,i\omega_m=\omega+i\delta)$ is the $c$ fermions spectral function. Since $\cosh(\beta \omega)^{-1}$ tends to unity for $\beta \omega \ll1$ and rapidly vanishes in the opposite limit $\beta \omega \gg 1$ , $G(k,\tau=\beta/2)$ amounts to an integral over the spectral function $A(k,\omega)$ over a frequency window of order $\sim T$. Further assuming a well-behaved spectral response for frequencies $\omega<T$ or equivalently no additional low energy excitations, we may use $\tilde{Z}(k)=\beta G(k,\tau=\beta/2)$ as an estimate for the $c$ electron single particle residue $Z=A^\beta(k,\omega=0)$.
	
	To detect the presence of Dirac fermions, we study the finite-size scaling properties of the superfluid stiffness, $\rho_s$, which captures the long-wavelength transverse current density $J_\mu(q,\omega_m)$ response to an external electro-magnetic gauge field $\mathcal{A}_\nu(q,\omega_m)$. Within linear response theory,  $J_\mu(q,\omega_m)=\Pi_{\mu \nu}(q,\omega_m) \mathcal{A}_\nu(q,\omega_m)$. Focusing on the $\nu=\mu=x$ component, the electromagnetic response function equals   \cite{Scalapino_1993}
		\begin{equation}
		\Pi_{xx}(q,\omega_m)=-\left[\av{-K_{x}} -\av{J_x(q,\omega_m) J_x(-q,-\omega_m)}\right].
		\end{equation}
		Here, $K_{x}$ is the kinetic energy density associated with $x$ oriented links and the second term is the current-current correlation function.
		In our specific case,
		\begin{align}
		\begin{split}
		K_{x}=-\sum_{r,\alpha} w \sigma^z_{r,x}f^\dagger_{r,\alpha} f_{r+\hat{x},\alpha}+t \tau^z_{r}f^\dagger_{r,\alpha} \tau^z_{r+\hat{x}}f_{r+\hat{x},\alpha}+h.c.,
		\end{split}
		\end{align}
		and the current operator at site $r$ along the $x$ direction is given by,
		\begin{align}
		\begin{split}
		J_{r,x}=-i\left[\sum_{\alpha} w \sigma^z_{r,x}f^\dagger_{r,\alpha} f_{r+\hat{x},\alpha}+t \tau^z_{r}f^\dagger_{r,\alpha} \tau^z_{r+\hat{x}}f_{r+\hat{x},\alpha}-h.c.\right].
		\end{split}
		\end{align}
		With the above definition, we can compute $\rho_s$ as the limit,
		\begin{equation}
		\rho_s = \lim_{q_y \to 0} \rho_s(q_y) =\lim_{q_y \to 0}\Pi_{xx}(q_x=0,q_y,\omega_m=0)
		\end{equation}
	
	It is convenient to express the superfluid stiffness in terms of the orbital magnetic susceptibility $\chi(q) = -{\rho_s(q)}/{q^2}$. The singularity associated with the Dirac node leads to a diverging magnetic response at low momenta $\chi(q)=-{g_v g_s v_f}/({16|q|})$ \cite{Ando_2009}, with $g_v$ ($g_s$) being the valley (spin) degeneracy and $v_f$ is the Dirac fermion velocity. Consequently, at low momenta, $\rho_s(q_y)\sim q_y \sim 1/L$, so that $\rho_s$ exhibits a slow decay that is inversely proportional to the system size. By contrast, a Fermi liquid admits a finite diamagnetic response at zero momentum (Landau diamagnetism), such that the expected behavior is $\rho_s(q_y)\sim q_y^2 \sim 1/L^2$, which vanishes more rapidly than before.

	To probe a potential instability towards an antiferromagnetic (AFM) order, we monitor the finite-momentum equal-time spin fluctuations $\chi_S(k)=\av{\left( \sum_{r}e^{ik\cdot r} S^z_{r}\right)^2}$, where $S^z_r=n^f_\uparrow-n^f_\downarrow$ is the $f$ electron spin polarization along the $z$ axis. From $\chi_S(k)$, we can compute the staggered magnetization $M_{\text{AFM}}=\sqrt{\chi_S(G_{\text{AFM}})/L^2}$, with $G_{\text{AFM}}=\{\pi,\pi\}$ being the Bragg vector associated with AFM order. The zero temperature AFM order parameter is obtained by taking the thermodynamic limit $\Delta_{\text{AFM}}=\lim_{L\to\infty} \lim_{\beta\to\infty} M_{\text{AFM}}(L,\beta)$. In practice, for a given system size, we monitored the convergence of the staggered magnetization toward its zero temperature value. Following that, we extrapolated the finite size data to the infinite system size value, using a polynomial fit in powers of $1/L$.
	
	In the presence of dynamical matter fields, determining whether the gauge sector is confined or deconfined is a particularly challenging task due to charge screening. Standard methods, relying on evaluating Wilson loops, no longer sharply distinguish between the two phases. Alternative methods based on extracting the topological contribution to the entanglement entropy \cite{Kitaev_2006,Levin_2006} and the Fredenhagen–Marcu order parameter \cite{Gregor_2011} are difficult to reliably scale with system size in fermionic systems. Instead, following refs.~\cite{GazitZ2,Gazit_pnas}, we probe the thermodynamic singularity associated with the confinement transition by tracking the Ising flux susceptibility $\chi_{B}={\partial \av{\Phi}}/{\partial K}$, a quantity that is expected to diverge at the confinement-deconfinement transition, akin to the specific heat singularity in classical phase transitions.
	
	% FIG - OSM to AFM
	\begin{figure}[!htb]
		\centering
		\subfp{subfig:flux_OSM_AFM} 
		\subfp{subfig:M_OSM_AFM}
		\includegraphics{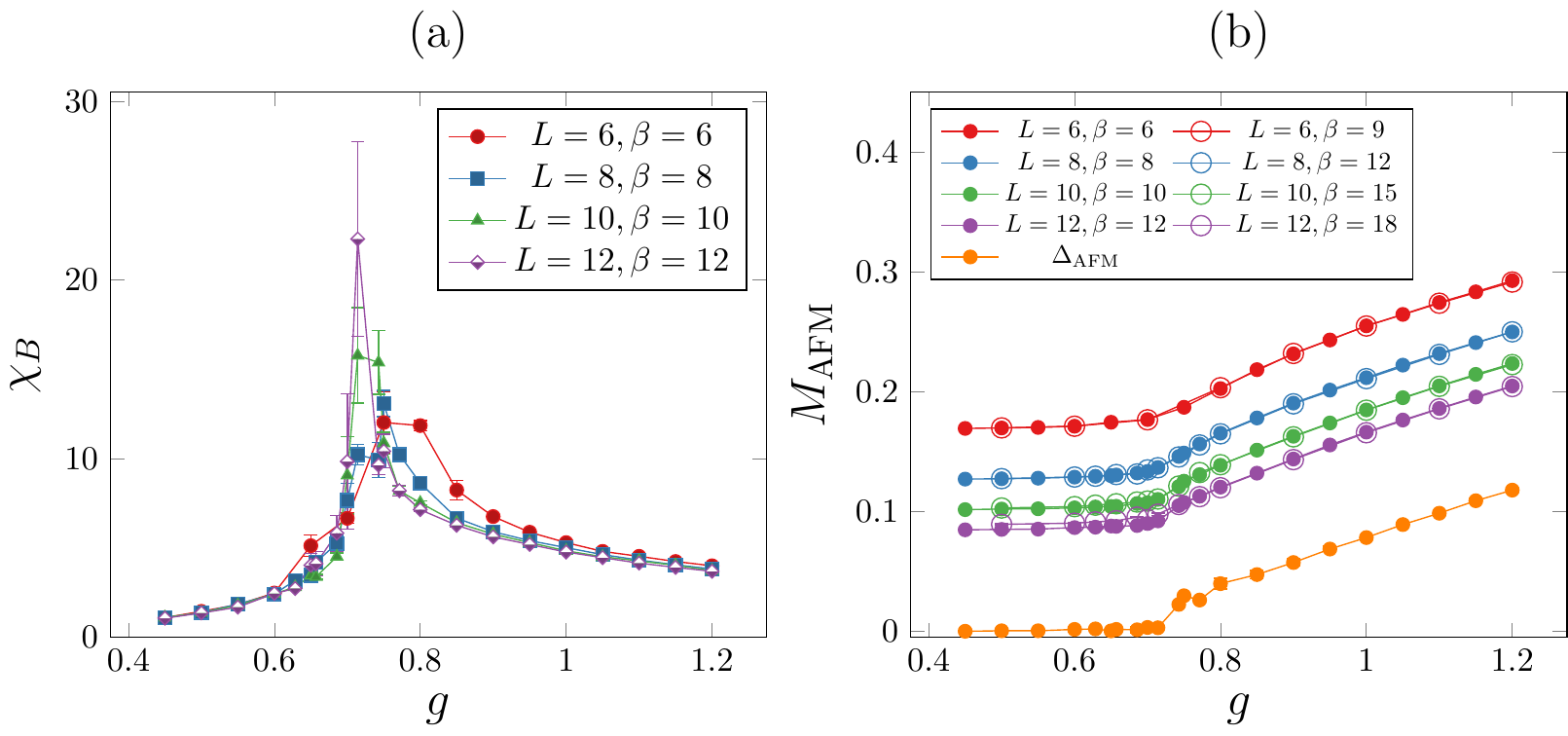}
		\caption{Orthogonal semi-metal confinement transition. Evolution of the (a) flux susceptibility $\chi_B$ and (b) staggered magnetization $M_{\mathrm{AFM}}$ across the phase transition separating the OSM and the confined AFM as a function of the transverse field $g$ for a fixed hopping amplitude $t=0.2$. Different curves correspond to a set of increasing system sizes and inverse temperatures. $\Delta_{\mathrm{AFM}}$ was obtained through a extrapolation of the finite size data to the thermodynamic limit. }
		\label{fig:conf_trans}
	\end{figure}
	
	\subsection{Numerical Results}
	
	% FIG -- OSM to FL spectral function 
	\begin{figure}[!htb]
		
		\subfp{subfig:ZK_OSM_to_FL_1} 
		\subfp{subfig:ZK_OSM_to_FL_2}
		\subfp{subfig:ZK_OSM_to_FL_3}
		\subfp{subfig:ZK_AFM_to_FL_1} 
		\subfp{subfig:ZK_AFM_to_FL_2}
		\subfp{subfig:ZK_AFM_to_FL_3}
		\includegraphics[scale=0.55]{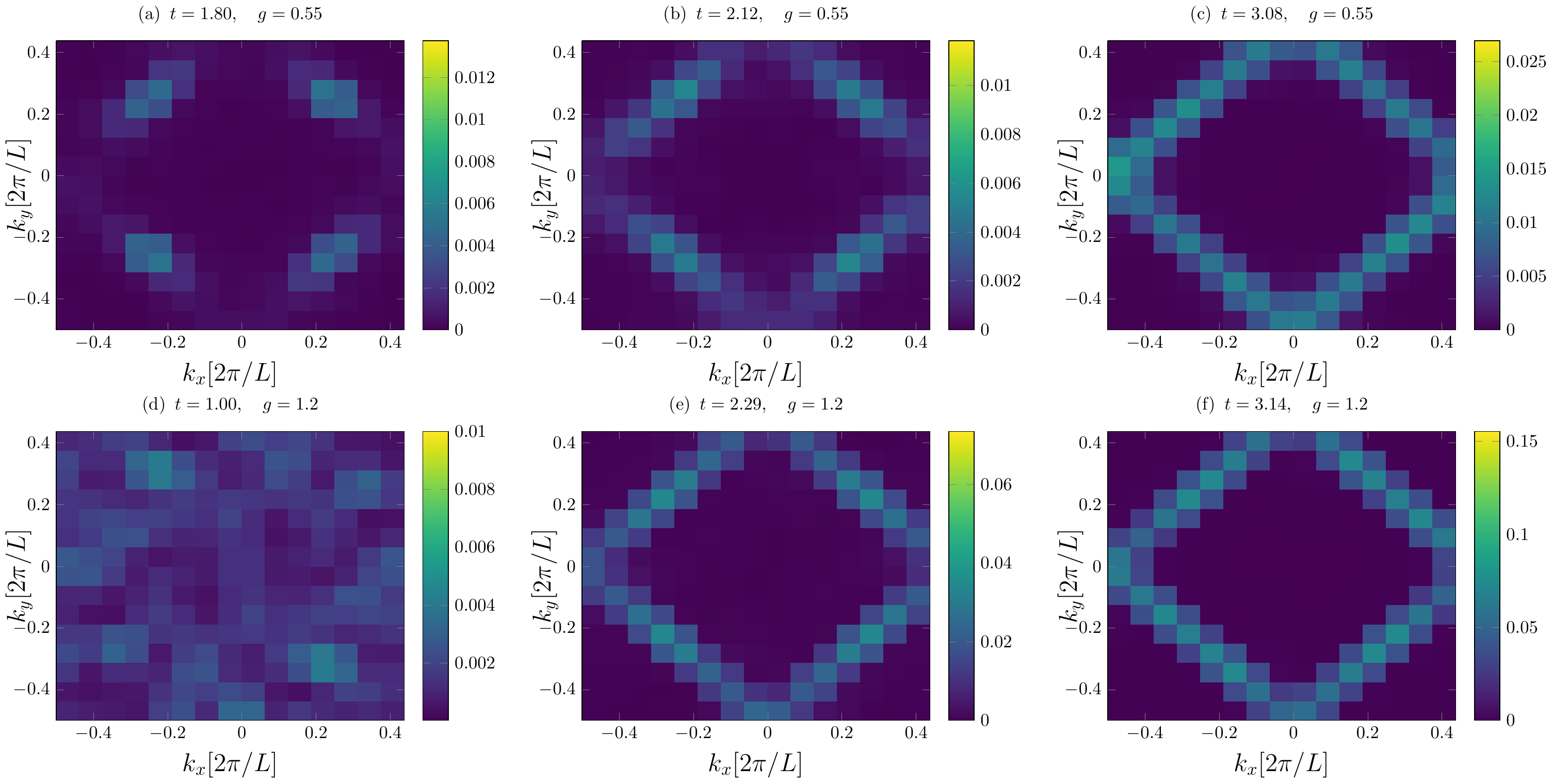}
		
		\caption{Momentum resolved $\tilde{Z}(k)$, for a linear system size $L=16$ and inverse temperature $\beta=16$, as a function the hopping amplitude $t$ along two parameter cuts: (i) In the top panels (a-c) we fix $g=0.55$ and cross the transition between the OSM and deconfined FL phases. Deep in the OSM phase we find four maxima with a finite spectral weight centered about $k=\{
			\pm\pi/2,\pm\pi/2\}$. with an increase in $t$, a large diamond shaped Fermi surface gradually appears upon approach to the deconfined FL phase (ii) In the bottom panels (d-e), on the other hand, we set $g=1.2$ and monitor the appearance of a Fermi surface, in the large $t$ limit, starting from a featureless low-energy spectrum deep in the confined AFM phase.  }
		\label{fig:ZK}
	\end{figure}
	
	% FIG -- OSM to FL anti-nodal spectral weight
	\begin{figure}[ht]
		\centering
		\begin{tabular}{cc}
			\subfp{subfig:Z_anti_nodal_OSM_FL} 
			\subfp{subfig:Lrhos} 
			\subfp{subfig:AFM_OSM_FL}
			\subfp{subfig:flux_OSM_FL} 
			\includegraphics[scale=0.9]{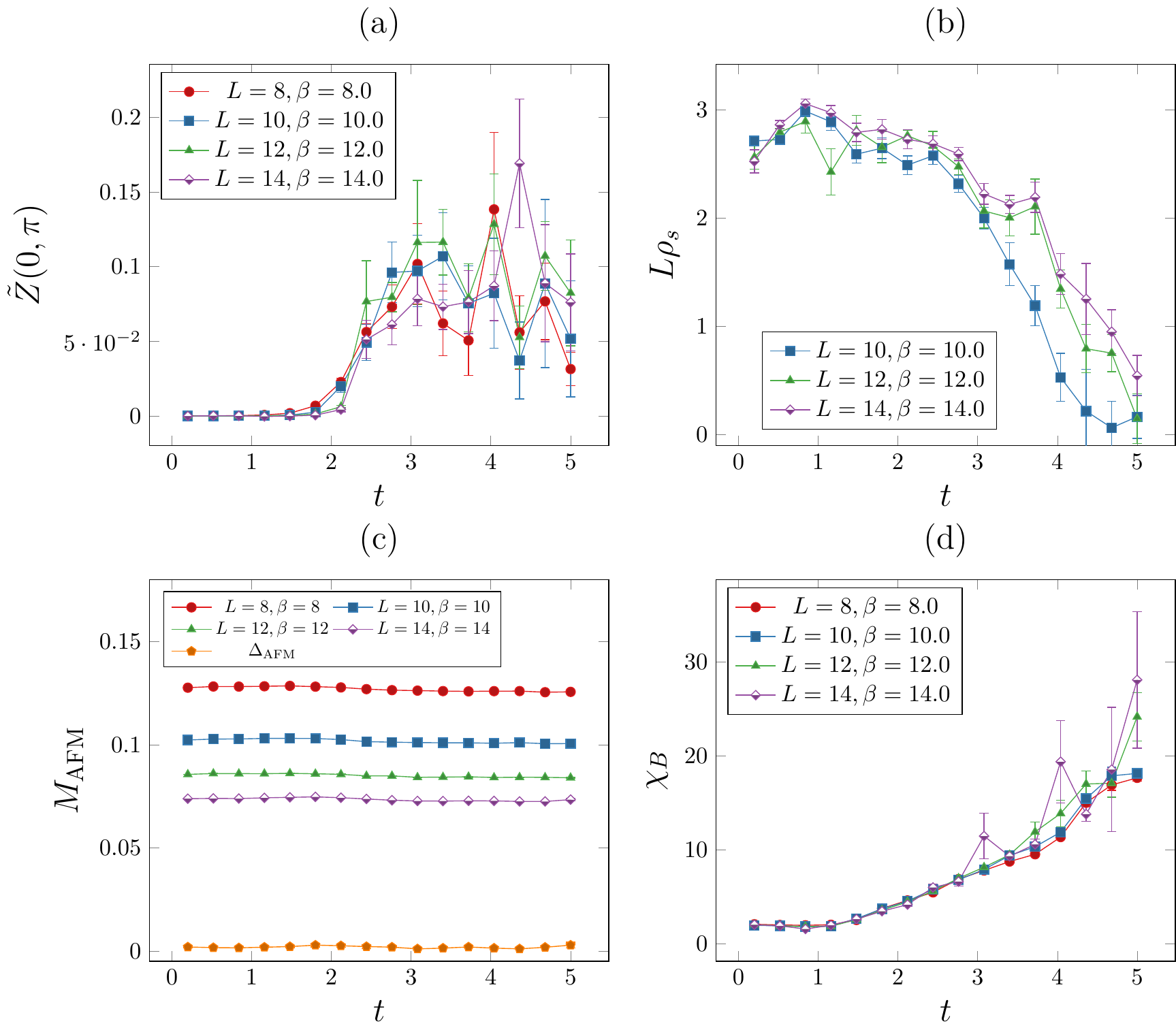}
		\end{tabular}
		\caption{Phase transition between the OSM and deconfined FL phases as a function of $t$ for $g=0.55$ (a) Single particle residue, $\tilde{Z}(k)$, at the anti-nodal points. (b) Critical finite size scaling of the super fluid stiffness, $L\rho_s$. (c) Staggered magnetization, $M_{\mathrm{AFM}}$. $\Delta_{\mathrm{AFM}}$ is computed by an extrapolation to the thermodynamic limit. (d) Flux susceptibility, $\chi_B$. } 
	\end{figure}
	
	To render the numerical computation tractable, we must restrict the relatively large parameter space spanned by the set of coupling constants appearing in $\mathcal{H}$. Starting from either the deconfined ($g\ll K$) or confined phase ($g\gg K$) our goal is to probe the emergence of physical $c$ fermions and the resulting formation of FL phases in the limit of large hopping amplitude, $t$. To that end, we numerically map out the phase diagram as a function of the transverse field, $g$, and physical fermion hopping amplitude, $t$. Throughout, we will consider a negative Ising flux coupling constant, $K<0$, for which a $\pi$-flux lattice is energetically favorable in deconfined phases.
	
	More concretely, we fix the microscopic parameters $w=-1,J=0.1,h=1.0,U=0.1$, and $K=-1$. All energy scales are measured in units of $|K|$. We note that we have chosen $h$ to be sufficiently large compared to $J$ in order to avoid condensation of Ising matter fields. The resulting two-parameter phase diagram is depicted in \cref{subfig:phasediag}.
	
	We begin our analysis by examining the limiting case $t\ll g$. In this regime, together with the above choice of microscopic parameters, the Ising matter field $\tau^z$ is gapped, and consequently also the physical fermion, $c$, is expected to be gapped. The low energy physics then involves only orthogonal fermions coupled to a fluctuating Ising gauge field. This physical setting was already studied extensively in previous works \cite{GazitZ2,Grover16,Gazit_pnas}. 
	
	In the context of our problem, we expect to find a similar structure of quantum phases in the above parameter regime: (i) A confining phase ($g
	\gg K$), where the orthogonal fermions together with the on-site background static charge (we consider an odd lattice gauge theory) form a localized gauge-neutral bound state, leaving the electronic spin as the only dynamical degree of freedom. Subsequently, quantum fluctuations will generate an effective antiferromagnetic Heisenberg coupling, leading to AFM order at zero temperature. (ii) In the deconfined phase ($g
	\ll K$), on the other hand, the orthogonal fermions are free and their dispersion is determined by the background flux configuration. For the case of a $\pi$-flux lattice, the band structure consists of two gapless and linearly dispersing bands.
	
	To numerically test the above reasoning, in \cref{fig:conf_trans}, we fix $t=0.2$ and plot the evolution of the flux susceptibility and staggered magnetization as a function of $g$. Indeed, in agreement with refs. \cite{Grover16,GazitZ2,Gazit_pnas}, we can identify the aforementioned phases: a deconfined OSM phase for small $g$, and with an increase in $g$, we observe a transition towards a confining phase accompanied with AFM order. We use a finite size scaling analysis to estimate the location of both the confinement and AFM symmetry breaking transitions. The flux susceptibility, see \cref{subfig:flux_OSM_AFM}, develops a peak at $g_c=0.75(5)$ that increases with system size and marks the position of the confinement transition. Concomitantly, in \cref{subfig:M_OSM_AFM}, we find that the AFM order parameter, $\Delta_{\mathrm{AFM}}$ begins to rise at $g_c=0.75(5)$.
	
	Due to the increased complexity of the model considered in this work, we found it challenging to reliably estimate universal data associated with the OSM confinement transition, such as critical exponents. Hence, we were unable to make a direct comparison with previous works. Nevertheless, the key signature of the OSM confinement transition, namely the non-trivial co-incidence of confinement and symmetry breaking \cite{Gazit_pnas} is fully consistent with the numerical data.
	
	\begin{figure}[ht]
		\centering
		\begin{tabular}{cc}
			
			\subfp{subfig:flux_FL_FL} 
			\subfp{subfig:anti_nodal} 
			\includegraphics{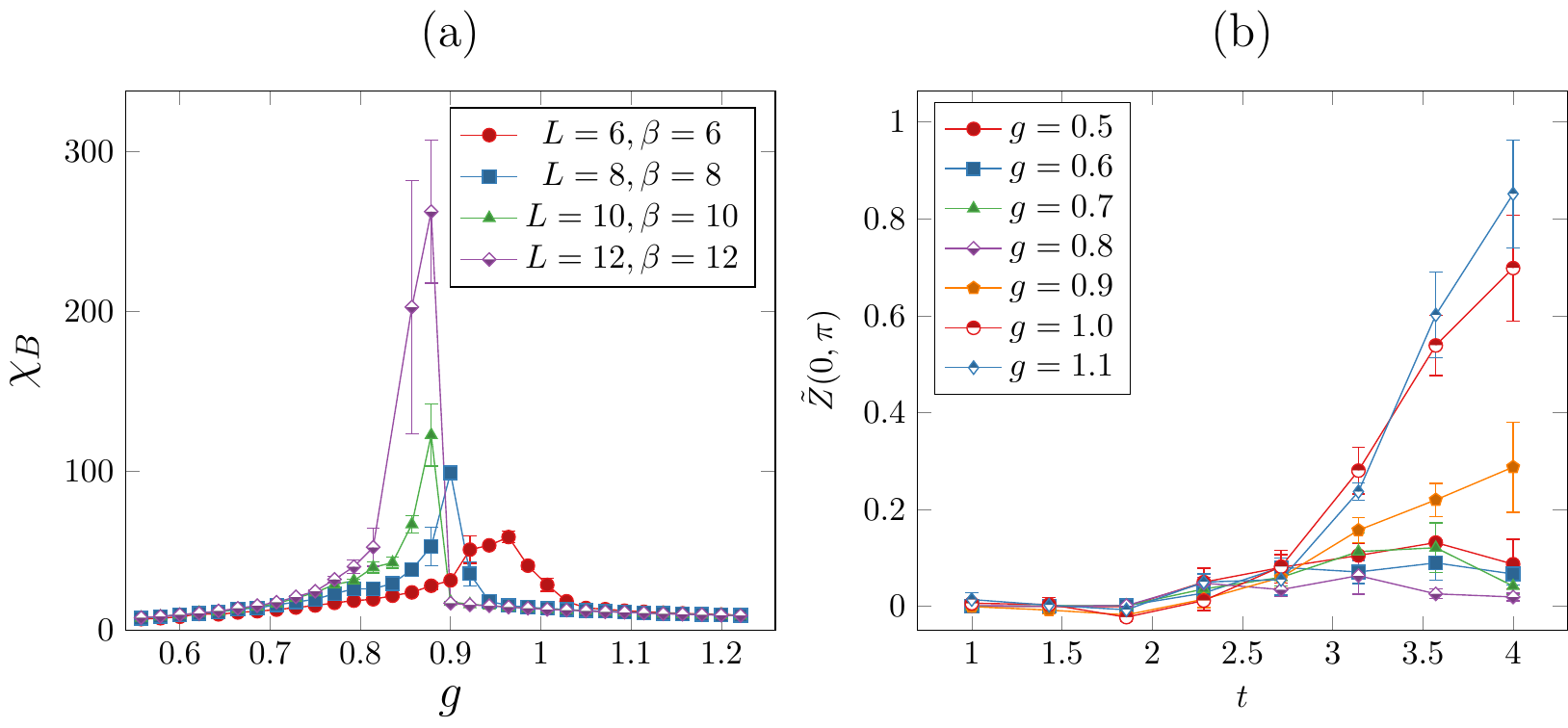}
		\end{tabular}
		\caption{(a)  Flux susceptibility $\chi_B$ as a function of $g$ for $t=3.0$  along the path connecting the confined and deconfined FL phases. (b) Single particle residue $\tilde{Z}(k)$ evaluated at the anti-nodal point $k=\{0,\pi\}$ as a function of $t$. Different curves correspond to different values of $g$  } 
	\end{figure}
	
	We now turn to address the main inquiry of this study, namely the emergence of low energy gauge-neutral $c$ fermions and their associated spectral signatures. With that goal in mind, in \cref{subfig:ZK_OSM_to_FL_1,subfig:ZK_OSM_to_FL_2,subfig:ZK_OSM_to_FL_3}, we depict our numerical estimate for the momentum resolved $c$ electron residue, $\tilde{Z}(k)$, at $g=0.55$ and for several increasing values of $t$, beginning from the OSM phase. Remarkably, we find that, in the OSM phase, $\tilde{Z}(k)$, comprises four maxima located at momenta $k=\{\pm\pi/2,\pm\pi/2\}$. This result is at odds with the conventional LT, which, at half-filling and in the absence of topological order or translational symmetry breaking, predicts a large Fermi surface encompassing half of the Brillouin zone. At large $t$ values, it is energetically favorable for the orthogonal fermion and $\tau$ particle to form a gauge-neutral bound state, identified with the $c$ electron, which effectively decouples from the gauge sector. Indeed, with an increase in $t$, the spectral function continuously evolves into the standard diamond shaped Fermi surface in compliance with LT.

	It is tempting to identify the observed ``Fermi pockets'' with the nodal points of the OSM. This explanation, however, is incorrect because the orthogonal $f$ fermion is not a gauge invariant object, and in particular the location of the Dirac nodes in momentum space is a gauge-dependent quantity. In \cref{sec:mf}, we provide a simple explanation for this phenomenon using a mean field calculation of the spectral function in the background of a static $\pi$-flux configuration.

	To further track the appearance of $c$ electrons, in Fig.~\ref{subfig:Z_anti_nodal_OSM_FL}, we set $g=0.55$ and study the evolution of the single particle residue evaluated at the anti-nodal point, $\tilde{Z}(0,\pi)$, along the path connecting the OSM and deconfined FL phases as a function of $t$.  Indeed, we observe that the spectral weight vanishes for $t<t_c\approx2.0$ and continuously rises for $t>t_c$, signaling the appearance of a large Fermi surface.  
	
	Next, we examine how the emergence of a finite spectral weight for the physical fermions $c$ influences the Dirac orthogonal fermions. To that end, in \cref{subfig:Lrhos} we examine the finite-size scaling behavior of the superfluid stiffness by plotting $L\rho_s$ as a function of $t$ for $g=0.55$. We find that for $t<t_c$, curves corresponding to different system sizes collapse to a single curve, namely the superfluid stiffness follows a critical scaling $\rho_s\sim 1/L$. As discussed before, this scaling behavior is characteristic of a low energy gapless Dirac spectrum. Unexpectedly, for $t>t_c$, a regime in which we previously found a finite quasiparticle weight at the anti-nodal point, we do not observe the expected Fermi-Liquid scaling $\rho_s\sim 1/L^2$, but rather a behavior consistent with $\rho_s\sim 1/L$. Although, we can not exclude the possibility that this behavior is due to a finite size crossover, our numerical results indicate the presence of an intermediate phase where orthogonal Dirac fermions and a FL of physical fermions coexist. In Section \ref{sec:qpt}, we provide a candidate theoretical description of this scenario.  Lastly, for $t \approx 5.0$, the product  $L \rho_s$ vanishes, signaling the absence of low energy orthogonal fermions. The parameter regime $t>5$ is beyond our current numerical capabilities. 
	
	To further examine the transition, we test for the development of AFM order or confinement in the gauge sector. In \cref{subfig:AFM_OSM_FL}, we  track the evolution of the staggered magnetization as a function of the hopping amplitude, $t$ (with $g=0.55$, as before). We do not find numerical evidence for a finite AFM order, even when a large Fermi surface is fully developed at large $t$. Moving to the gauge sector, in \cref{subfig:flux_OSM_FL}, we probe $\chi_B$  along the same trajectory as above. The flux susceptibility appears to cross the transition smoothly. We can, therefore, conclude that the Fermi surface reconstruction involves neither translational symmetry breaking nor the loss of topological order. Thus, the phase at large $t$ is a deconfined FL.
	
	We remark that due to the perfect nesting condition of the half-filled square lattice, the ground state is expected to exhibits AFM order for arbitrarily small Hubbard interaction. This phenomenon was not observed in our simulations, since in the weak coupling regime $U\ll t$ the magnetization is exponentially small in the coupling constant, and its detection is a notoriously difficult numerical task. 
	
	Next, we examine the path connecting the confined AFM state and the FL phase, by setting $g=1.2$ and probing the evolution of the spectral function as a function of $t$. The results of this analysis are shown in \cref{subfig:ZK_AFM_to_FL_1,subfig:ZK_AFM_to_FL_2,subfig:ZK_AFM_to_FL_3}. We observe a featureless flat spectrum deep in the confined AFM phase, as expected due to the absence of fermionic quasiparticles. By contrast, with an increase in $t$, a large Fermi surface appears. 
	
	To better appreciate the above result, we note that in a confining phase, the low energy spectrum must contain solely gauge-neutral excitations. Indeed, in the confined AFM phase, we can identify these excitations with spin-waves. However, fractionalized orthogonal fermions, carrying $U(1)$ electromagnetic charge, are localized. On the other hand, with an increase in $t$ the attractive force between $\tau^z$ and $f$ allows forming a gauge-neutral bound states of $c$ fermions and a metallic state that supports both spin {\em and} charge excitations at low energies.

	We now turn to study the confinement transition along a path connecting the deconfined and confined FL phases. To detect the confinement transition, in \cref{subfig:flux_FL_FL}, we plot the flux susceptibility at $t=3.0$ as a function of $g$. Indeed, we observe a divergence in $\chi_B$ at $g_c=0.85(5)$ marking the location of the confinement transition. The above result demonstrates that our model sustains FL phases in the background of either a confined or a deconfined gauge sector.
	
	Lastly, we summarize our spectral analysis in \cref{subfig:anti_nodal}, where we plot the spectral weight, $\tilde{Z}(0,\pi)$, evaluated on the ``anti-nodal" point as a function of the hopping amplitude $t$ for several values of $g$. As a starting point, at low $t$, We consider both the OSM and  AFM phases.  We find that for all $g$ values the spectral weight is small at low $t$ and rises continuously starting from  the energy scale $t_c(g)$. Operationally, we numerically estimate $t_c(g)$ by locating the hopping amplitude for which $\tilde{Z}(0,\pi)>0.01$. This analysis was used to mark the phase boundaries appearing in \cref{subfig:phasediag}. We remark again, that at half-filling the Fermi-liquid phase is unstable towards the formation of AFM order. Consequently, at strictly zero temperature $\tilde{Z}(0,\pi)$ must vanish for all $t$ and the transition between the AFM and Fermi liquid phases is a smooth cross-over.

	% FIG -- Confined AFM to FL spectral function 

	% FIG - AFM - g cuts

	\subsection{Quantum phase transitions}
	\label{sec:qpt}
	
	We now briefly remark on the theoretical expectations for the quantum phase transitions described above and appearing in Fig.~\ref{subfig:phasediag}.
	\begin{itemize}
		\item The theory for the transition from the OSM to the AFM was discussed in some detail in ref.~\cite{Gazit_pnas}, and identified as a deconfined critical point with an emergent SO(5) symmetry.
		\item Away from half-filling, the AFM to Confined FL transition is a conventional symmetry-breaking transition between two confining phases, and is expected to be described by Landau-Ginzburg-Wilson theory combining with damping from Fermi surface excitations, {\it i.e.\/} Hertz-Millis theory \cite{ssbook}.
		\item The transition from the Deconfined FL to the Confined FL phase is a confinement transition without a change in the size of the Fermi surface. It is therefore expected to described by the condensation of an Ising scalar, which can be viewed as representing either the `vison' of the Ising gauge theory or the Ising matter field $\tau^z$ \cite{SenthilFisher}. The large Fermi surface of electron-like quasiparticles will damp the quasiparticles, but this damping is much weaker than that in Hertz-Millis theory: the resulting field theory was described in refs.~\cite{SachdevMorinari,Bilocal}. We note that this field theory also applies to the confinement transition in ref.~\cite{Meng3}, where the damping is due to a large Fermi surface of orthogonal fermions, and there is no change in the size of the Fermi surface across the transition.
		\item 
		We do not have a theory for a direct transition from the OSM to the Deconfined FL with a large Fermi surface transition. Indeed,  it may well be that this transition occurs via an intermediate phase, with gapless excitations of both electrons and orthogonal fermions \cite{Powell05,Kaul08}. The presence of a regime within the ``Deconfined FL'' region of the phase diagram where we appear to observe co-existence of a $c$ Fermi surface (as measured by the $c$ spectral density) with a $\rho_s$ which vanishes as $\sim 1/L$ is evidence in support of such an intermediate phase. The Luttinger constraint requires that the orthogonal fermions which are absorbed into the $c$ Fermi surface must be accounted for by those forming the Dirac nodes. One way to preserve the Dirac nodes in an intermediate phase, and also maintain consistency with particle-hole symmetry, is to form electron-like and hole-like pockets of an equal area of the $c$ fermions; the $c$ hole pockets would then be in ancillary trivial insulator, similar to refs.~\cite{Yahui1,Yahui2}. Our resolution is not sharp enough to resolve such an intricate Fermi surface evolution, which is likely present in the intermediate phase, and we leave its study to future work.
	\end{itemize}
	\begin{figure}[th!]
		\centering
		\begin{tabular}{cc}
			
			\begin{subfigure}[b]{0.5\textwidth}
				\centering
				\includegraphics[width=0.5\textwidth]{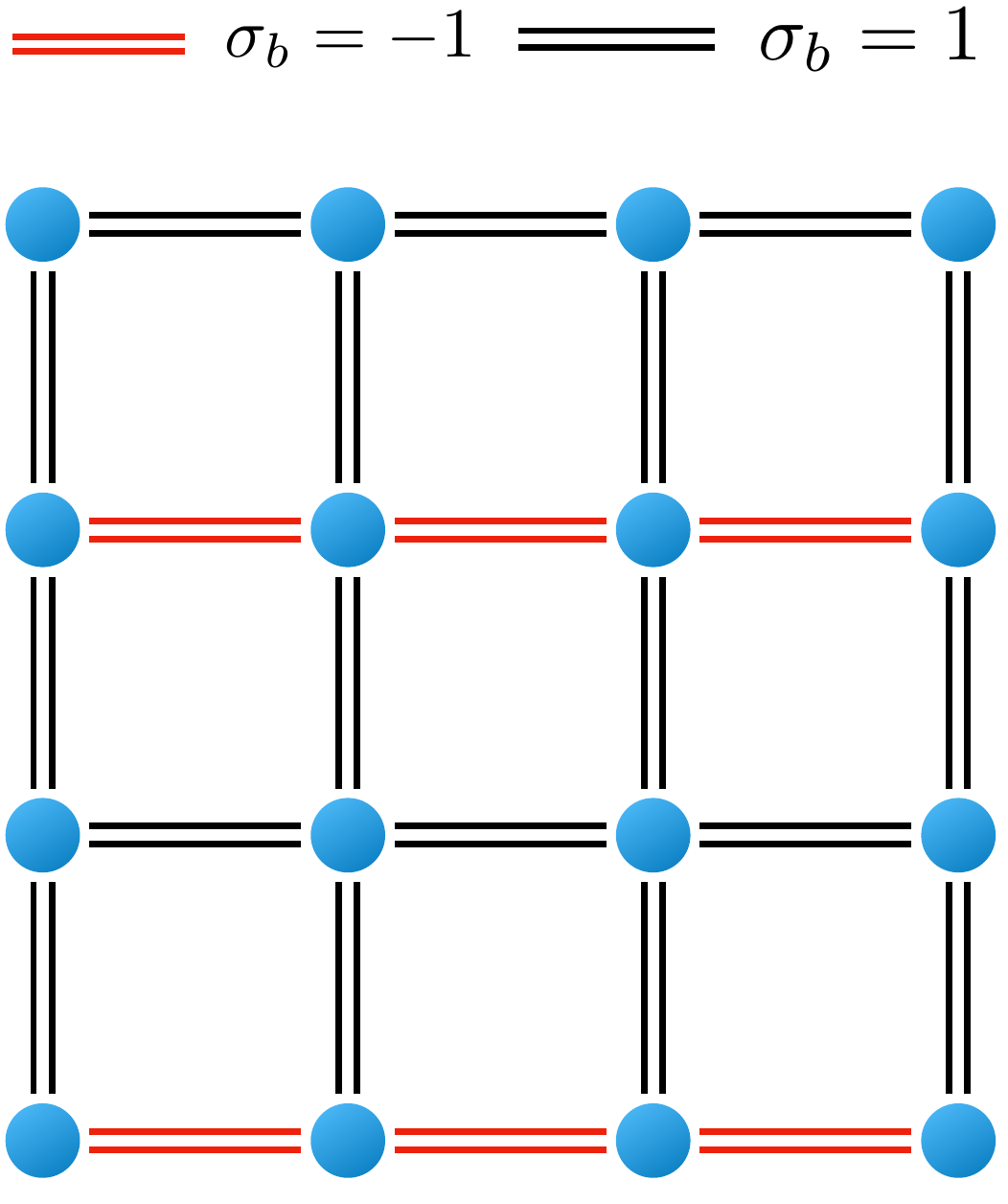}
				
				\caption{}
				\label{subfig:pi_flux_lattice}
			\end{subfigure}&
			\begin{subfigure}[b]{0.5\textwidth}
				\centering
				\includegraphics[width=1.0\textwidth]{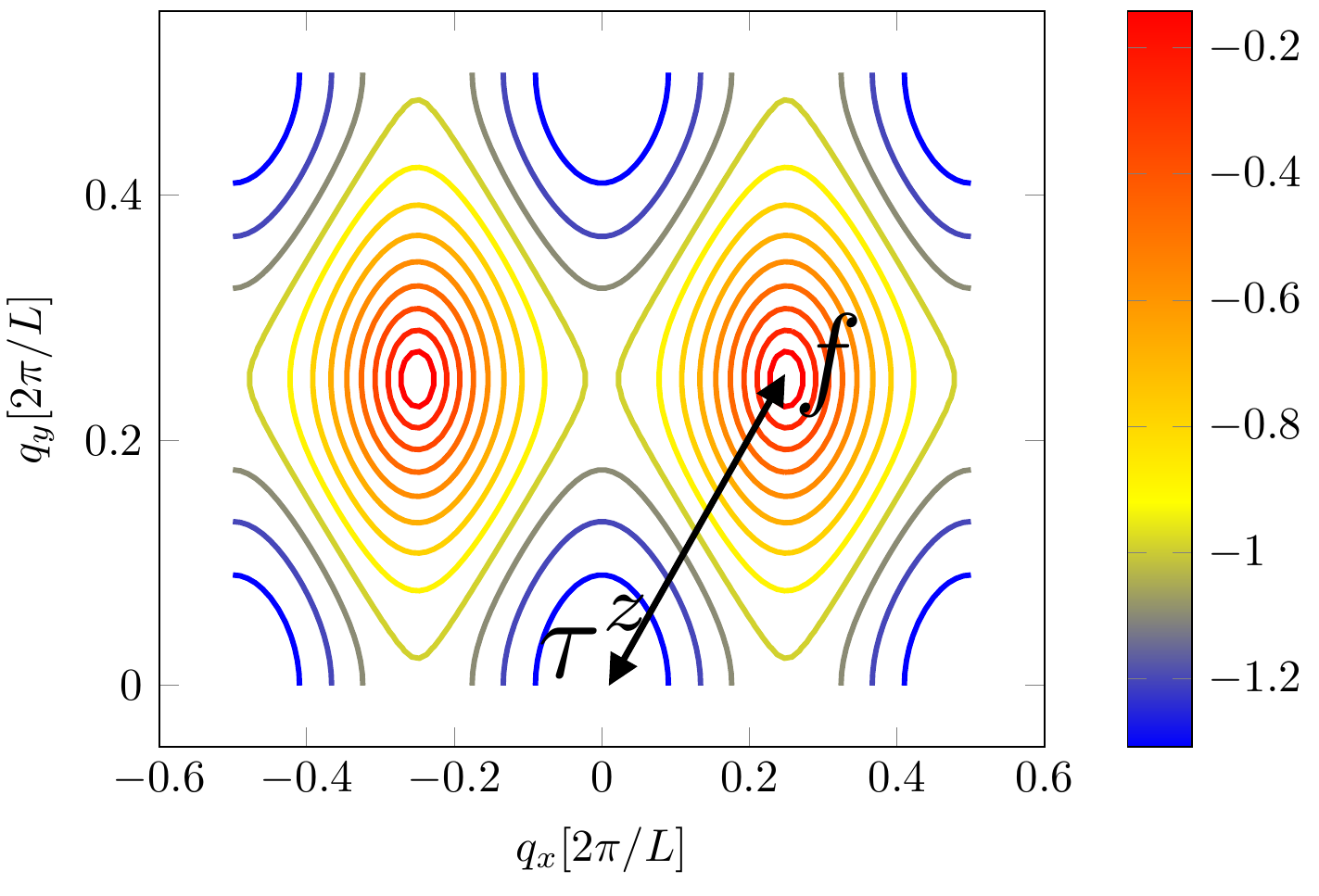}
				\caption{}
				\label{subfig:dirac}
			\end{subfigure} 
			\\
			\begin{subfigure}[b]{0.5\textwidth}
				\centering

			\raisebox{15mm}{	\includegraphics[width=0.9\textwidth]{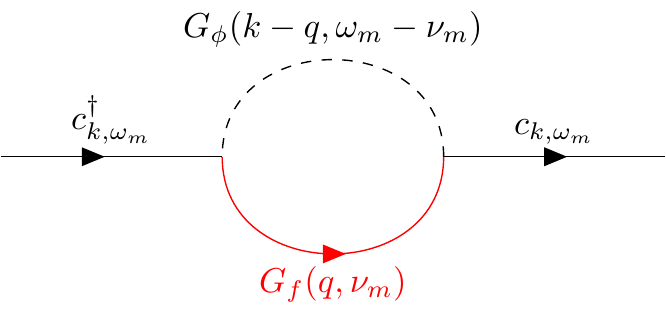}}
							\caption{}
				\label{subfig:bubble_diag}
			\end{subfigure} 
			&
			\begin{subfigure}[b]{0.5\textwidth}
				\centering
				\hspace*{-2em}\includegraphics[width=1.0\textwidth]{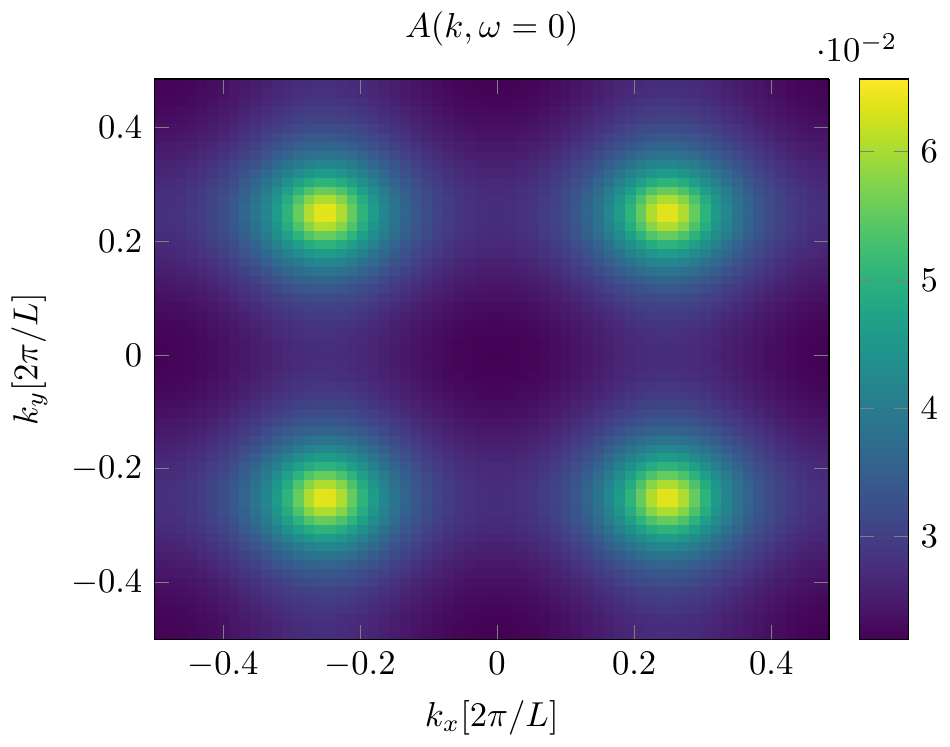}
				\caption{}
				\label{subfig:Akw}
			\end{subfigure}
		\end{tabular}
		\caption{(a) Ising Landau gauge fixing condition for the $\pi$-flux lattice, black (red) bonds corresponds to $\sigma^z_b=1(-1)$ (b) Energy contours of the lower band Dirac spectrum on the $\pi$-flux lattice using the above gauge fixing. (c) Leading order Feynman diagram for the physical fermion $c$ propagator $\mathcal{G}_c^\beta(k,i\omega_m)$. The bubble diagram evaluates to a convolution between the Ising matter field $\phi^z$ and `orthogonal' fermion $f_\sigma$ propagators defined on the $\pi$-flux lattice (d) Finite temperature spectral function $A(k,\omega=0)$ computed by a numerical evaluation of the bubble diagram.  }
	\end{figure}
	\section{Mean field calculation of the physical fermion spectral function in the OSM phase.}
	\label{sec:mf}
	
	The numerical observation of a finite spectral weight centered about the four nodal points $k=\{\pm\pi/2,\pm\pi/2\}$ is surprising and requires further analytic understanding. To that end, in this section, we will present a simple and intuitive explanation using a mean-field calculation of the spectral function of $c$ fermions in the OSM phase. In our calculation, we consider a static background $\pi$-flux configuration and neglect gauge field fluctuations. This approximation is justified deep in the OSM phase, where such fluctuations are small due to the finite vison gap. In this setting, we can model the dynamics of the $f$ electron ( $\tau^z$ field) by a free fermion (scalar field) hopping in the background of a $\pi$-flux lattice.
	
	To make progress, we must choose a concrete a gauge fixing condition for the Ising gauge field. We take an Ising variant of Landau's gauge, $\sigma_{r,\hat{x}}=(-1)^{r_y}$ and $\sigma_{r,\hat{y}}=1$, see \cref{subfig:pi_flux_lattice}. We emphasize that while the gauge fixing procedure inevitably breaks both lattice translations and $\pi/2$ rotations, expectation values of physical (gauge-neutral) observable must preserve these symmetries. More concretely, we take the effective Hamiltonians,
	\begin{equation}
	\begin{aligned}
	\mathcal{H}^{\text{MF}}_f&=-t_f\sum_{r,\eta} t_{r,\eta} f_{r}^\dagger f_{r+\eta} +h.c.\\
	\mathcal{H}^{\text{MF}}_\phi&=\sum_{r} \frac{\pi_r^2}{2m} +\frac{m \omega^2}{2}\left( \sum_r \Delta \phi_{r}^2 +\sum_{r,\eta}(\phi_r-t_{r,\eta}\phi_{r+\eta})^2\right),
	\end{aligned}
	\label{eq:eff_H}
	\end{equation}
	
	Here, $\phi_r$ is a scalar real field and $\pi_r$ is its canonically conjugate momentum, $[\phi_r,\pi_{r'}]=i\delta _{r,r'}$. With the above gauge choice the hopping amplitudes are set to $t_{r,\eta}=\left((-1)^{r_y}\delta_{\eta,\hat{x}}+\delta_{\eta,\hat{y}}\right)$. The scalar field Hamiltonian is parameterized by the inertial mass $m$, oscillation frequency $\omega$, and $\Delta$, which allows controlling the single particle excitation gap for $\phi$ particles.
	
	We tune $\Delta$ to work in a regime where, on the one hand, the scalar field has a finite gap to avoid condensation, but, on the other hand, it is sufficiently small to render the physical fermion gap small. The lowest order diagram \cite{Podolsky_2009} contributing to $\mathcal{G}^\beta_c(k,i\omega_m)$ is the bubble diagram shown in ~\cref{subfig:bubble_diag}. Evaluating the diagram boils down to a convolution of the Ising matter field propagator, $\mathcal{G}^\beta_{\phi}(k,i\omega_m)$, and the orthogonal fermion propagator, $\mathcal{G}^\beta_f(k,i\omega_m)$. Both propagators can be readily computed by diagonalizing the quadratic Hamiltonians in ~\cref{eq:eff_H}. Further details of this calculation are given in \cref{app:bub_diag_eval}.
	
	In \cref{subfig:Akw} we depict the zero-frequency spectral function $A(k,\omega=0)=-\frac{1}{\pi}\Imag\left[\mathcal{G}^\beta_c(k,i\omega_m=i0^+)\right]$, evaluated for the microscopic parameters $t=m=\omega=1$ and $\Delta=-1.1$, and inverse temperature $\beta=1/32$.
	Remarkably, our simplified model displays a finite spectral weight located at the nodal points $k=\{\pm\pi/2,\pm\pi/2\}$, in agreement with the exact QMC calculation. As a non-trivial check of our computation, we observe that unlike the gauge fixed hopping Hamiltonians of \cref{eq:eff_H}, the physical spectral function respects the {\em full} square lattice $C_4$ point group symmetry. We note that at strictly zero temperature, the spectral weight must vanish due to the non-zero Ising matter field gap.
	
	An intuitive understanding of this result may be derived by examining the Dirac band structure in the $\pi$-flux phase, shown as a contour plot in ~\cref{subfig:dirac}. At low temperatures, the orthogonal fermions will occupy all $k$-space modes up to the Dirac point at $\{\pm \pi/2,\pi/2\}$. However, the Ising matter field, $\tau^z$ follows Bose statistics and hence will concentrate at the band minimum, $\{0/\pi,0\}$. As a result, the integral over the internal momentum in \cref{subfig:bubble_diag} will be appreciable only at momentum transfer $k=\{\pm\pi/2,\pm\pi/2\}$ (black arrow in \cref{subfig:dirac} connecting the $f$ and $\tau^z$ particle). In other words, the momentum-space splitting of the two flavors of matter fields is responsible for the unconventional spectral response of the OSM phase.

	\section{Discussion and summary}
	\label{sec:summary}
	
	We have studied a lattice model of orthogonal-fermions coupled to an Ising-Higgs gauge theory. The absence of the sign-problem enabled us to determine its global phase diagram and explore related phase transitions using a numerically exact QMC simulation. A key ingredient of our study, which non-trivially distinguishes it from previous works, is the introduction of an Ising matter field that together with the orthogonal fermion may form a physical fermion. This crucial feature of our model enabled access to the study of FL phases in the presence of $\ztwo$ topological order.
	
	Notably, our model hosts both non-FL quantum states that violate LT due to $\ztwo$ topological order, and also LT preserving FL states, where the gauge sector is either confined or deconfined. On tuning of microscopic parameters, we are able to cross quantum critical points that separate these phases, some of which appear to be continuous.
	
	It is interesting to make a connection, even if suggestive, between our numerical results and experimental signatures of Fermi surface reconstruction observed in cuprate materials. In particular, the OSM phase shares several properties with the pseudo-gap phase: a strong depletion in the density of states at the anti-nodal points, and concentration of finite fermionic spectral weight at the nodal points $k=\{\pm\pi/2,\pm/2\}$. Most importantly, there is experimental evidence that these phenomena occur in the absence of translational symmetry breaking. We also described phase transitions involving the appearance of a large Fermi surface from such a state, and this has connections to phenomena in the cuprates near optimal doping \cite{Keimer2015,PTARCMP,Fujita14,He14,badoux_change_2016,MichonNature,Ramshaw2020,Sachdev_2018}.

	Our study has left some open questions. In particular, it would be interesting to develop a field theory description to the transition between the OSM and deconfined FL, which numerically appears to be continuous. Such a description will have to address the unusual finite-size scaling of the superfluid stiffness in the deconfined FL phase.  From the numerical perspective, simulations on larger lattices are key in resolving the properties of this transition. In addition, it would be interesting to study the fate of the OSM phase and neighboring phases away from half-filling. This can be achieved, using a sign-problem free QMC, at least for the even lattice gauge theory case. We leave these interesting questions for future studies.

	{\it A note added:} During the final stages of this work, we have become aware of a related work by Chen {\it et al.\/} \cite{Meng2} studying a lattice model of orthogonal fermions in a parameter regime complementary to our work, which we described briefly in \cref{sec:intro} and \cref{sec:qpt}.
	
	\begin{acknowledgments}
		We thank Ashvin Vishwanath, Aavishkar A. Patel, and Amit Keren for useful discussions. S.G. acknowledges
		support from the Israel Science Foundation, Grant No. 1686/18.
		S.S. was supported by the National Science Foundation under Grant No. DMR-1664842. S.S. was also supported by the Simons Collaboration on Ultra-Quantum Matter, which is a grant from the Simons Foundation (651440, S.S.)
		F.F.A. thanks the DFG collaborative research centre SFB1170 ToCoTronics (project C01) for financial support as well as the W\"{u}rzburg-Dresden Cluster of Excellence on Complexity and Topology in Quantum Matter – ct.qmat (EXC 2147, project-id 390858490).
		This work was partially performed at the Kavli Institute for Theoretical Physics (NSF grant PHY-1748958). FFA gratefully acknowledges the Gauss Centre for Supercomputing e.V. (www.gauss-centre.eu) for funding this project by providing computing time on the GCS Supercomputer SUPERMUC-NG at the Leibniz Supercomputing Centre (www.lrz.de). This research used the Lawrencium computational cluster resource provided by the IT Division at the Lawrence Berkeley National Laboratory (Supported by the Director, Office of Science, Office of Basic Energy Sciences, of the U.S. Department of Energy under Contract No. DE-AC02-05CH11231) and the Intel Labs Academic Compute Environment.
	\end{acknowledgments}
	
	\appendix
	
	\section{Path integral formulation}
	\label{app:path_integral}
	The partition function, $\mathcal{Z}(\beta)$ at inverse temperature $\beta$ is given by,
	\begin{equation}
	\mathcal{Z}(\beta)=Tr\left[\hat{P}e^{-\beta \mathcal{H}}\right]
	\end{equation}
	where the projection operator $\hat{P}=\prod_iP_i$ with,
	\begin{equation}
	\hat{P}_i=\frac{1}{2}\left(1+(-1)^{n^f_i}\tau^x_i\prod_{+}\sigma_{ij}^x\right)
	\end{equation} 
	We rewrite the projection operator using a discrete Lagrange multiplier as,
	\begin{equation}
	\hat{P}_i=\sum_{\lambda_i\pm1}\hat{P}_{i,\lambda}=\sum_{\lambda_i=\pm1}e^{i\frac \pi 2 (1-\lambda_i)\left(\sum_{+}\left(\frac{1-\sigma^x_{i,j}}{2}\right)+\frac{1-\tau_i^x}{2} +n^f_i\right)}
	\end{equation}
	
	Next, we use Trotter decomposition to write $e^{-\beta \mathcal{H}}=\prod_{m=0}^{M-1} e^{-\epsilon\mathcal{H}}$ with $\epsilon=\beta/M$ and insert a resolution of identities in the $\sigma^z$ and $\tau^z$ basis, $\mathds{1}=\sum \ket{\tau^z,\sigma^z}\bra{\tau^z,\sigma^z}$ leading to,
	\begin{equation}
	\mathcal{Z}(\beta)=\sum_{\lambda,\tau^z,\sigma^z}Tr_f \bra{\tau^z_{0},\sigma^z_{0}} \prod_i\hat{P}_{i,\lambda} e^{-\epsilon\mathcal{H}}\ket{\tau^z_{M-1},\sigma^z_{M-1}}\ldots \bra{\tau^z_{1},\sigma^z_{1}}  e^{-\epsilon\mathcal{H}}\ket{\tau^z_{0},\sigma^z_{0}}
	\end{equation}
	
	We first focus on the last time step, which contains $\hat{P}_{i,\lambda}$. In the Ising sector, the only off-diagonal (in the $\sigma^z,\tau^z$ basis ) terms are the transverse field and the constraint. Focusing on a specific site $i$, for the $\ztwo$ matter field we obtain the matrix element,
	\begin{equation}
	\begin{aligned}
	\bra{\tau^z_{i,0}}  e^{i\frac \pi 2 (1-\lambda_i)\left(\frac{1-\tau_i^x}{2} \right)+\epsilon h \tau_i^x}\ket{\tau^z_{i,M-1}}&=\sum_{\tau^x_i=\pm1}\bra{\tau^z_{i,0}}  e^{i\frac \pi 2 (1-\lambda_i)\left(\frac{1-\tau_i^x}{2} \right)+\epsilon h \tau_i^x}\ket{\tau^x_{i}}\bra{\tau^x_{i}}\ket{\tau^z_{i,M-1}}\\
	&=\frac{1}{2}\sum_{\tau^x_i=\pm1} e^{\epsilon h \tau_i^x} e^{i\pi\left(\frac{1-\tau^x_i}{2}\right)\left(\frac{1-\lambda_i}{2}+\frac{1-\tau^z_{i,0}}{2}+\frac{1-\tau^z_{i,M-1}}{2}\right)}\\
	&=\frac{1}{2}\left(e^{\epsilon h}+e^{-\epsilon h}e^{i\pi\left(\frac{1-\lambda_i}{2}+\frac{1-\tau^z_{i,0}}{2}+\frac{1-\tau^z_{i,M-1}}{2}\right)}\right)\\
	&=\begin{cases}
	\cosh(\epsilon h) & \lambda_i\tau^z_{i,0} \tau^z_{i,M-1}=1 \\
	\sinh(\epsilon h) &  \lambda_i\tau^z_{i,0} \tau^z_{i,M-1}=-1 
	\end{cases}
	\end{aligned}
	\end{equation}
	The effective Boltzmann weight is then,
	\begin{equation}
	W (\lambda_i,\tau^z_{i,0} ,\tau^z_{i,M-1})\propto e^{\gamma\tau^z_{i,0} \lambda_i \tau^z_{i,M-1} }
	\end{equation}
	where $\gamma=-\frac{1}{2}\log(\tanh(\epsilon h))$.
	Physically, the above action corresponds to the gauge invariant Ising interaction along the temporal direction.
	Importantly, we must take $h>0$ in order to avoid a sign problem.
	
	Similarly to ref.\cite{GazitZ2}, the constraint term associated with the Ising gauge field leads to a spatio-temporal plaquette term in the 3D Ising gauge theory, and the $f$ fermions Green's function is modified by the introduction of a diagonal matrix $P[\lambda_i]$ with diagonal elements $P_{ii}=\lambda_i$.

	\clearpage
	\section{Details of the mean field calculation}
	\label{app:bub_diag_eval}
	
	To evaluate the bubble diagram in \cref{subfig:bubble_diag}, we first express the mean field Hamiltonians (\cref{eq:H_ising,eq:Hf}) in momentum space defined on a reduced Brillouin zone ($0<k_x<2\pi,0<k_y<\pi$), as imposed by the gauge fixing choice in \cref{subfig:pi_flux_lattice}. Explicitly, for the fermionic part, substituting $f_r=\frac{1}{\sqrt{N}}\sum_{k}e^{ikr}f_k$ ($N$ being the number of lattice sites) gives
	
	\begin{equation}
	\begin{aligned}
	\mathcal{H}^{\mathrm{MF}}_f&=-t_f\sum_{k,k'} f^\dagger_{k}f_{k'}\left(2\cos(k_y)\delta_{k,k'}+2\cos(k_x)\delta_{k,k'+\pi\hat{k}_y}\right)\\
	&= -t_f\sum_{0<k_x<2\pi,0<k_y<\pi} \begin{pmatrix} f^\dagger_{k,0} & f^\dagger_{k,\pi} \end{pmatrix}
	\begin{pmatrix} 2\cos{k_y} &  2\cos{k_x} \\ 2\cos{k_x} &  -2\cos{k_y}\end{pmatrix}
	\begin{pmatrix} f_{k,0} \\ f_{k,\pi} \end{pmatrix}
	\end{aligned}
	\end{equation}
	where $f^\dagger_{k,0}=f^\dagger_{k}$ and $f^\dagger_{k,\pi}=f^\dagger_{k+\pi\hat{k}_y}$. Diagonalizing the above Hamiltonian, we obtain the fermionic eigen spectrum $\epsilon_\pm(k)$ and eigen-modes $f_\alpha(k)=V_{\alpha,\gamma}(k)f_\gamma(k)$, where $\gamma=\pm$ and $\alpha=0/\pi$ and $V_{\alpha,\gamma}$ is the diagonalizng matrix.
	
	A similar analysis is carried out in order to diagonalize the scalar field $\phi$ mean field Hamiltonian. Writing \cref{eq:eff_H}, in momentum space gives, ($\phi_r/\pi_r=\frac{1}{\sqrt{N}}\sum_k\phi_k/\pi_k e^{ikr}$)
	\begin{equation}
	\begin{aligned}
	\mathcal{H}_\phi&=\sum_{k} \frac{\pi_k \pi_{-k}}{2} +\frac{ m\omega^2}{2}\left( \sum_{k} \Delta \phi_k \phi_{-k}+\sum_{k,k'}\phi_{k}\phi_{k'} \left( 4 - 2 \cos(k_y) \delta_{k,-k'}-2\cos(k_x)\delta_{k,-k'+\pi \hat{k}_y}\right)\right) \\
	&=\sum_{0<k_x<2\pi,0<k_y<\pi} \sum_\alpha\frac{\pi_{k,\alpha} \pi_{-k,\alpha}}{2} +\frac{ m\omega^2}{2}\sum_{\alpha,\alpha'}\phi_{k,\alpha}K_{\alpha,\alpha'}(k)\phi_{-k,\alpha'},
	\end{aligned}
	\end{equation}
	where the spring constant matrix equals,
	\begin{equation}
	K_{\alpha,\alpha'}(k)=\begin{pmatrix} \Delta +4-2 \cos(k_y) &  -2\cos(k_x) \\  -2\cos(k_x) & \Delta +4+2 \cos(k_y) \end{pmatrix}.
	\end{equation}
	
	By diagonalizaing $K_{\alpha,\alpha'}$ (the mass matrix is already diagonal) we obtain the eigen-frequencies $\omega_{\kappa}(k)$ and the normal modes $\phi_{k,\alpha}=U_{\alpha,\kappa}(k)\phi_{k,\kappa}$. As before, $\kappa=\pm$, and $\alpha=0,\pi$.
	With the above definitions, our final expression of the bubble diagram amplitude reads,
	\begin{equation}
	\begin{aligned}
	\mathcal{G}_c(k,i\omega_m) &= \sum_{q,q',\nu_m,\gamma,\kappa} \delta_{p(q),p(q')} \delta_{p(k-q),p(k-q')}V_{\alpha(q),\gamma}(p(q)) \\
	&\times V_{\alpha(q'),\gamma}(p(q')) U_{\alpha(k-q),\kappa}(p(k-q)) U_{\alpha(k-q'),\kappa}(p(k-q')) \\ &\times \frac{1}{i\nu_m-\epsilon_\gamma(p(q))}  \frac{1}{((\nu_m-\omega_m)^2+\omega^2_\kappa(p(k-q))}
	\end{aligned}
	\end{equation}
	
	Here, the reduced momentum $p(q)$ is defined as $p(q)=\{q_x,q_y \mod \pi\}$ and the momentum index $\alpha(q)$ equals $0\,(\pi)$ if $q_y\in[0,\pi]$ ($q_y\in[\pi,2\pi]$), as before. We note that the resulting momentum integration is restricted to the two cases $q=q'$ or $q=q'+\pi\hat{q}_y$.

	The Matsubara sum can be evaluated analytically,
	\begin{equation}
	\sum_{\nu_m}\frac{1}{i\nu_m-\epsilon_\gamma}  \frac{1}{(\nu_m-\omega_m)^2+\omega^2_\kappa}=\frac{- \beta\omega _\kappa \tanh \left(\frac{\beta  \epsilon _\gamma}{2}\right)+ \beta\left(\epsilon_\gamma-i \omega_m\right) \coth \left(\frac{\beta  \omega_\kappa}{2}\right)}{2 \omega _\kappa\left(\omega_\kappa^2- \left(\epsilon _\gamma-i \omega_m\right)^2\right)}
	\end{equation}
	
	and we are left with the momentum integration that is computed numerically on a discretized momentum grid.

	\bibliography{FL}
\end{document}